# Parameterized Quantum Circuits with Quantum Kernels for Machine Learning: A Hybrid Quantum-Classical Approach

Daniel T. Chang (张遵)

*IBM (Retired)* [dtchang43@gmail.com](mailto:dtchang43@gmail.com)

**Abstract:** Quantum machine learning (QML) is the use of quantum computing for the computation of machine learning algorithms. With the prevalence and importance of classical data, a hybrid quantum-classical approach to QML is called for. Parameterized Quantum Circuits (PQCs), and particularly Quantum Kernel PQCs, are generally used in the hybrid approach to QML. In this paper we discuss some important aspects of PQCs with quantum kernels including PQCs, quantum kernels, quantum kernels with quantum advantage, and the trainability of quantum kernels. We conclude that quantum kernels with hybrid kernel methods, a.k.a. quantum kernel methods, offer distinct advantages as a hybrid approach to QML. Not only do they apply to Noisy Intermediate-Scale Quantum (NISQ) devices, but they also can be used to solve all types of machine learning problems including regression, classification, clustering, and dimension reduction. Furthermore, beyond quantum utility, quantum advantage can be attained if the quantum kernels, i.e., the quantum feature encodings, are classically intractable.

## 1 Introduction

*Quantum machine learning (QML)* is the use of *quantum computing* for the computation of *machine learning algorithms*, or part thereof. Our focus is on QML for *classical data*, as opposed to quantum data (information), since it is prevalent and it is used in standard machine learning problems. The expectation is that quantum processing will make it possible to considerably *accelerate* machine learning processes or, more importantly, tackle machine learning problems which are *classically intractable* due to computational time or cost.

With our focus on classical data and the fact that quantum information cannot be accessed without measurement, a *hybrid quantum-classical approach to QML* is called for [1]:

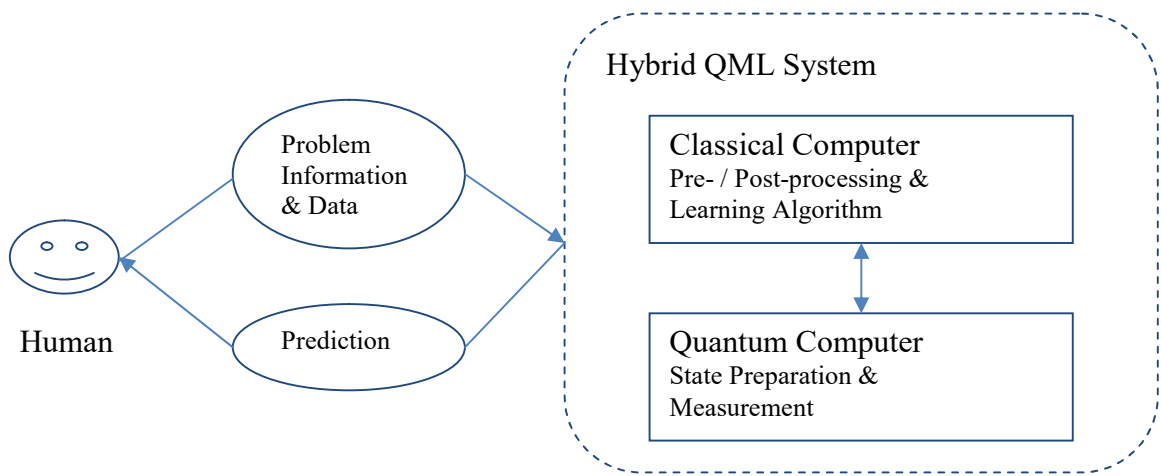

In the hybrid approach, the data is *pre-processed* and a set of *model parameters* is determined on a classical computer. The quantum computer prepares a *quantum state* as prescribed by a *quantum circuit* and performs *measurements*. Measurement outcomes are *post-processed* by the classical computer to generate a *prediction*. To improve the prediction, the classical computer implements a *learning algorithm* that updates the model parameters. The overall algorithm is run in a *closed loop* between the classical and quantum computer.

*Parameterized Quantum Circuits (PQCs)* [1-2] are generally used in the hybrid approach to QML. There are three types of QML models based on PQCs [2]: *Standard, Data Re-Uploading*, and *Quantum Kernel*. The Standard PQC is composed of fixed gates, as a *quantum feature encoding circuit*, and adjustable gates, as a *variational quantum circuit*. Even at low circuit depth, some classes of PQCs are capable of generating highly non-trivial outputs. Therefore, they offer a concrete way to implement QML algorithms on *Noisy Intermediate-Scale Quantum (NISQ) devices*. Furthermore, PQCs provide some of the most promising practical applications of QML to achieve *quantum advantage*.

*Kernel methods* [6-7] for machine learning are ubiquitous for pattern recognition. They have many successful practical applications, the most famous being the SVM, and they have a rich theoretical foundation. However, there are *limitations* to the success of kernel methods when the *feature space* becomes large, and the *kernel functions* become computationally expensive to estimate.

Kernel methods use positive semidefinite, symmetric kernel functions to encode data points into a Reproducing Kernel Hilbert Space (RKHS). *Quantum kernel* [3-5] naturally arises from the similarity between the RKHS and the Hilbert space of the quantum system. Quantum kernels can provide speedups [8-9, 12] or equal performance [13] in the exploitation of an *exponentially large quantum state space* through controllable entanglement and interference. Furthermore, quantum kernels can be readily used in (classical) kernel methods. This is a great advantage in a *hybrid approach to QML* as it allows PQCs to focus only on computing (classically-intractable) kernels and leverages the rest of the methods as is.

It is an important challenge to find quantum kernels with *quantum advantage*, which are *classically intractable*, for real-world data. A class of quantum kernels with quantum advantage, called *covariant quantum kernels*, that can be used for *group-structured data* is introduced in [9]. Machine learning problems with group-structured data have important practical applications. However, advanced classical kernels for group-structured data are known to be computationally expensive to evaluate and are mostly approximated. A potential advantage of quantum kernels, therefore, can be that such an approximation may not be necessary.

It is commonly believed that the *optimal quantum kernel-based model* can always be obtained [5] due to the convexity of the problem. This is true provided that the kernel values can be efficiently obtained to a sufficiently high precision. However, there exist scenarios where quantum kernels are *exponentially concentrated* towards some fixed value and so *exponential resources* are required to accurately estimate the kernel values. Three sources are identified in [11] that can lead to exponential concentration and untrainability for quantum kernels including: the *expressibility of data encoding*, *global measurements*, and *noise*.

In this paper we discuss in detail these important aspects of PQCs with quantum kernels, namely PQCs, quantum kernels, quantum kernels with quantum advantage, and the trainability of quantum kernels.

# 2 Parameterized Quantum Circuits for Machine Learning

*Parameterized Quantum Circuits (PQCs)* [1-2] are generally used in the hybrid approach to QML. There are three types of QML models based on PQCs [2]: *Standard, Data Re-Uploading*, and *Quantum Kernel*. The Standard PQC is composed of fixed gates, as a *quantum feature encoding circuit*, and adjustable gates, as a *variational quantum circuit*, as shown below:

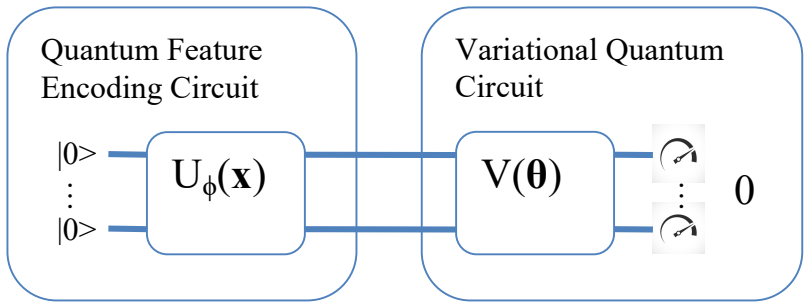

The Data Re-Uploading PQC is organized as a *series* of alternating feature encoding circuit and variational circuit, as shown below with L units:

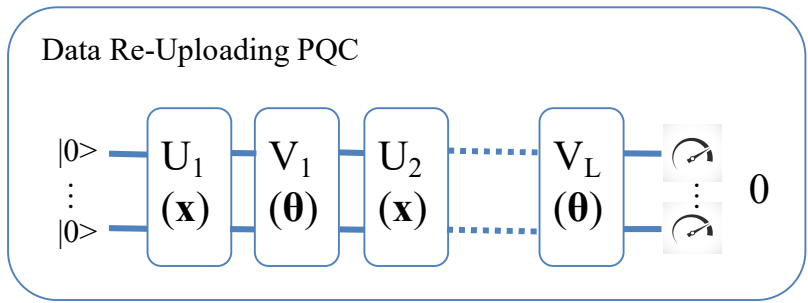

In the *Quantum Kernel PQC* the variational quantum circuit is not used. Instead, the *adjoint* of the quantum feature encoding circuit is used to compute the *inner product* of two quantum states, which defines a *quantum kernel* (as discussed in Section 3 Quantum Kernels for Machine Learning):

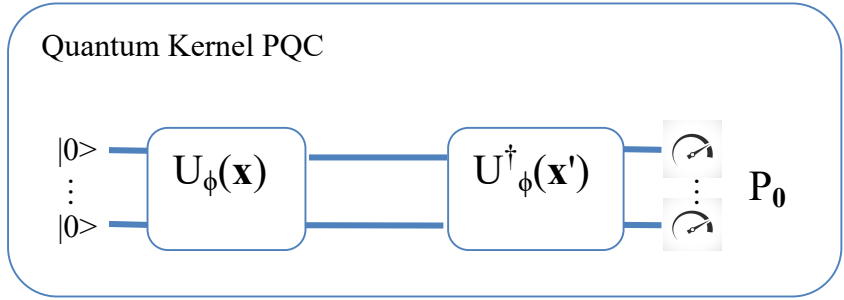

Note that the the quantum feature encoding circuit may be *parameterized* (for an example, see Section 4.1 Covariant Quantum Kernels).

The Standard PQC and the Data Re-uploading PQC are *explicit models* which use the quantum computer to directly learn a *linear decision boundary* in feature space by optimizing a variational quantum circuit. The Quantum Kernel PQC, on the other hand, is an *implicit model* which takes a classical model that depends on a *kernel function*, but uses the quantum computer to evaluate the kernel. Our focus is on the *Quantum Kernel PQC*.

Even at low circuit depth, some classes of PQCs are capable of generating highly non-trivial outputs. Therefore, they offer a concrete way to implement QML algorithms on *NISQ devices*. Furthermore, PQCs provide some of the most promising practical applications of QML to achieve *quantum advantage*.

## 3 Quantum Kernels for Machine Learning

*Kernel methods* [6-7] for machine learning are ubiquitous for pattern recognition, with support vector machines (SVMs) and Gaussian processes being the most well-known methods. They use a distance measure, a.k.a. *kernel function*, $k(x, x')$ between any two data points **x** and **x'** in order to construct models that capture the properties of a data distribution. This kernel function is connected to *inner products* in the *feature space*. Kernel methods have many successful practical applications, the most famous being the SVM, and they have a rich theoretical foundation. However, there are *limitations* to the success of kernel methods when the *feature space* becomes large, and the *kernel functions* become computationally expensive to estimate.

Kernel methods use *positive semidefinite, symmetric* kernel functions to encode data points into a *Reproducing Kernel Hilbert Space (RKHS)*. *Quantum kernel* [3-5] naturally arises from the similarity between the RKHS and the *Hilbert space of the quantum system*. Quantum kernels can provide speedups [8-9, 12] or equal performance [13] in the exploitation of an *exponentially large quantum state space* through controllable entanglement and interference.

With quantum kernels the process of encoding *data x* into a quantum state is interpreted as a *nonlinear feature map $\phi(x)$* [4] which maps the data into a potentially vastly higher-dimensional feature space, the Hilbert space of the quantum system. Data can then be analyzed in this *feature Hilbert space*. Furthermore, the *inner product* of two data points that have been mapped into the feature Hilbert space gives rise to a *kernel function*. Specifically, the *quantum feature map* [1, 3-4]

$$\mathbf{x} \rightarrow U_\phi(\mathbf{x})|0^n\rangle$$

represents a mapping to the high-dimensional vector space of the states of *n qubits*. The inner product of two data points in this space defines a *quantum kernel function* [1, 3]

$$k(\mathbf{x}, \mathbf{x}') = |\langle 0^n| U^\dagger_\phi(\mathbf{x}') U_\phi(\mathbf{x}) |0^n\rangle|^2.$$

Note that linear transformations are natural for quantum computing, but nonlinear transformations on the data are difficult to design in quantum computing. The *nonlinear feature map* approach `out-sources' [4] the nonlinearity into the procedure of encoding data into a quantum state and therefore offers an elegant solution to the problem of nonlinearities.

Resorting to the representer theorem, the *model function* of kernel methods is expressed as an expansion over kernel functions [1]

$$f(\mathbf{x}, \mathbf{w}) = \sum_{i=1}^{N} w_i k(\mathbf{x}, \mathbf{x}^{(i)}).$$

The learning task is to find *parameters w* so that the model outputs correct predictions. Note that these parameters define the *classical post-processing function*, as opposed to an operation of the PQC.

Quantum kernels can be readily used in (*classical*) *kernel methods* such as the SVM, the Gaussian process, and the principal component analysis. This is a great advantage in a *hybrid approach to QML* as it allows PQCs to focus only on computing (classically-intractable) kernels and leverages the rest of the methods as is. As such, QML algorithms optimized with data can fundamentally be formulated as *hybrid kernel methods* [5] whose *kernel* is computed by a quantum computer but the *rest of the method* is processed by a classical computer. This means that while the *quantum kernel* itself may explore high-dimensional state spaces of the quantum system, the *kernel method* can be trained and operated in low-dimensional spaces of the classical system.

## 3.1 Quantum Kernel Estimation (QKE)

Quantum kernels can be computed using the *Quantum Kernel PQC* with the *Quantum Kernel Estimation (QKE)* [3, 8] algorithm. The main idea [8] is to *map* classical data vectors into quantum states:

$$\mathbf{x} \rightarrow |\phi(\mathbf{x})\rangle\langle\phi(\mathbf{x})|,$$

where the *density matrix* representation is used to avoid global phase. Then, the *kernel function* is the Hilbert-Schmidt inner product between density matrices,

$$k(\mathbf{x}, \mathbf{x}') = \mathrm{Tr}[|\phi(\mathbf{x}')\rangle\langle\phi(\mathbf{x}')| \cdot |\phi(\mathbf{x})\rangle\langle\phi(\mathbf{x})|] = |\langle\phi(\mathbf{x}')|\phi(\mathbf{x})\rangle|^2.$$

This *quantum feature map* is implemented via a quantum circuit 'parameterized' by $\mathbf{x}$,

$$|\phi(\mathbf{x})\rangle = U_\phi(\mathbf{x})|0^n\rangle,$$

where it is assumed the feature map uses *n qubits*. Therefore, to compute the *quantum kernel function*

$$k(\mathbf{x}, \mathbf{x}') = |\langle 0^n|U^\dagger_\phi(\mathbf{x}')U_\phi(\mathbf{x})|0^n\rangle|^2,$$

we can run the Quantum Kernel PQC, $U^\dagger_\phi(\mathbf{x}')U_\phi(\mathbf{x})$, on input $|0^n\rangle$, and *measure the probability of the $0^n$ output* with a sufficient number of measurements [3].

## 4 Quantum Kernels with Quantum Advantage

It is an important challenge to find quantum kernels with *quantum advantage*, which are *classically intractable*, for *real-world data*. Learning algorithms that use QKE have proven advantage over all classical learners for specifically constructed data [8]. However, the core challenge is to establish quantum advantage for real-world data.

A class of quantum kernels, with quantum advantage, that can be used for *group-structured data* is introduced in [9]. The kernel is defined in terms of a unitary representation of the group and a *fiducial state* that can be optimized using a technique called *quantum kernel alignment (QKA)*. Specifically, it considers a general class of quantum feature encoding circuits, called *covariant feature maps,* which can be used for group-structured data, and uses them to estimate *covariant quantum kernels*. The details are discussed in the first two subsections that follow.

Machine learning problems with *group-structured data* have important practical applications. However, advanced classical kernels for group-structured data are known to be computationally expensive to evaluate and are mostly approximated. A potential advantage of *quantum kernels*, therefore, can be that such an approximation may not be necessary.

### 4.1 Covariant Quantum Kernels

The *covariant feature map* is defined relative to a *unitary representation* $D_x$ for the *group G* with $x \, \varepsilon \, G$, and a *fiducial state* $|\psi\rangle$ *on n qubits* as

$$\Phi(x) = D_x|\psi\rangle\langle\psi|D^{\dagger}_x.$$

The covariant quantum kernel is then estimated as

$$k(x, x') = |\langle\psi|D^{\dagger}_{x'}D_x|\psi\rangle|^2.$$

The *fiducial state* can be prepared by applying an efficient *quantum circuit V*:

$$|\psi\rangle = V|0^n\rangle.$$

The QKE routine for the *covariant quantum kernel* then reduces to estimating the transition amplitude

$$k(x, x') = |\langle 0^n|V^{\dagger}D^{\dagger}_{x'}D_xV|0^n\rangle|^2,$$

and the *quantum feature encoding circuit* becomes

$$U_{\phi}(x) = D_xV,$$

as shown in the *Covariant Quantum Kernel PQC* (with a variational quantum circuit $V_{\lambda}$):

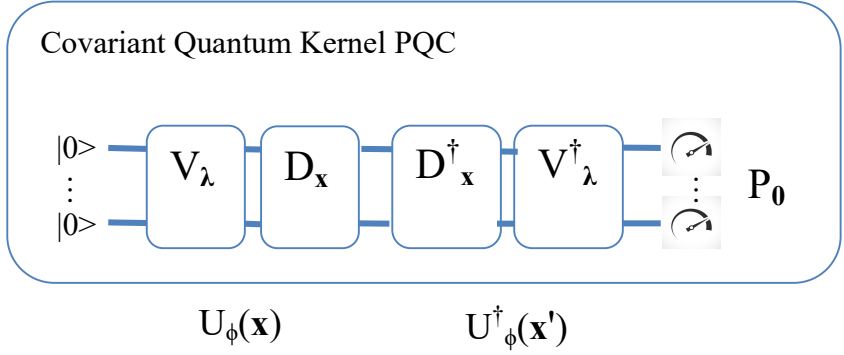

The covariant quantum kernel as defined above is *left-invariant* under the group action. The *right-invariant* version is

$$k(\mathbf{x}, \mathbf{x}') = |\langle 0^n|V^\dagger D_\mathbf{x} \cdot D^\dagger_{\mathbf{x}'} V|0^n\rangle|^2.$$

### 4.2 Quantum Kernel Alignment (QKA)

*Covariant quantum kernels* can lead to a *provable separation* between quantum and classical learners for specific problems. The choice of the *fiducial state* is essential for the performance of the covariant quantum kernel. If sufficient structural knowledge about the problem is present, a suitable fiducial state can be *chosen a priori*. If no prior knowledge is available, the fiducial state can be *optimized subject to the available data*. The objective of this optimization will depend on the *learning problem*.

In the latter case, the parameters $\lambda$ of the variational quantum circuit $V_\lambda$ can be optimized with *kernel alignment*. For *binary classification* problems, the *model function* associated with kernel $k_\lambda(\mathbf{x}, \mathbf{x}')$ is given as a linear threshold function

$$f(\mathbf{x}) = \text{sign}(\sum_{i=1}^m \alpha_i y_i k_\lambda(\mathbf{x}, \mathbf{x}_i))$$

with model parameters $\alpha_i$ for a training set of size m and labels $y_i = \pm 1$. The *SVM* is used as the kernel method to optimize the model parameters with the cost function

$$F(\boldsymbol{\alpha}, \lambda) = \sum_{i=1}^m \alpha_i - (1/2)\sum_{i,j=1}^m \alpha_i \alpha_j y_i y_j k_\lambda(\mathbf{x}_i, \mathbf{x}_j),$$

which is an upper bound to the generalization error when *maximized* over $\boldsymbol{\alpha}$. The *weighted* kernel alignment *minimizes* this upper bound with respect to $\lambda$. The procedure is thus expressed as the optimization,

$$\min_\lambda \max_\alpha F(\alpha, \lambda).$$

Note that the *weighted kernel alignment* aims at finding a kernel using the given data that also maximizes the gap margin of the SVM classifier. This means that only the training points that are *support vectors* contribute towards learning the kernel.

A *gradient-descent based stochastic algorithm* is used for this optimization problem, which is an iterative algorithm with *kernel matrices* evaluated on a quantum computer and *parameters updated* with classical optimization routines.

### 4.3 Software Package and Quantum Advantage Seeker for Quantum Kernels

IBM Qiskit provides basic facilities for *designing and experimenting quantum kernels* [3] on quantum simulators and IBM Quantum Computers. It provides Python interfaces including QuantumKernel, QuantumKernelTrainer, QuantumKernelTrainerResult, KernelLoss, ZZFeatureMap, QSVC, and SVCLoss. Furthermore, it provides tutorials with code examples on quantum kernel estimation (QKE), quantum kernel alignment (QKA), and a quantum kernel training toolkit.

The process of *designing, experimenting, and evaluating different quantum kernels* can be nontrivial. *QuASK (Quantum Advantage Seeker with Kernel)* [10] is a software package written in Python that supports designing, experimenting, and evaluating different quantum and classical kernels performance. It can be integrated with *IBM Qiskit*, among other quantum software packages. QuASK guides the user through a simple *preprocessing of input data*, definition and calculation of *quantum and classical kernels*. From this evaluation, it provides an assessment about potential *quantum advantage* and prediction bounds on *generalization error*. Furthermore, QuASK can generate the observable values of a quantum model and use them to study the *prediction capabilities* of the quantum and classical kernels. QuASK is promising. However, there are no tutorials with Qiskit code examples to demonstrate its use or capabilities.

### 5 Trainability of Quantum Kernels

It is commonly believed that the *optimal quantum kernel-based model* can always be obtained [5] due to the *convexity* of the problem. This is true provided that the kernel values can be efficiently obtained to a sufficiently high precision. However, there exist scenarios where quantum kernels are *exponentially concentrated* towards some fixed value and so *exponential resources* are required to accurately estimate the kernel values. This is due to the fact that it can be extremely difficult to extract any useful information from the *exponentially large Hilbert space*, especially in the presence of noise.

Three sources are identified in [11] that can lead to exponential concentration and untrainability for quantum kernels including: the *expressibility of data encoding*, *global measurements*, and *noise*. For each source, an associated concentration bound of quantum kernels is analytically derived. Furthermore, it is shown that when dealing with *classical data*, training a quantum feature encoding circuit with a *QKA* algorithm, as discussed in Section 4.2 Quantum Kernel Alignment (QKA), is also susceptible to exponential concentration. The three sources that can lead to exponential concentration are discussed in the following subsections.

## 5.1 Expressibility-induced Concentration

For a *quantum feature encoding circuit* $U_\phi(x)$ an ensemble of unitaries can be generated over all possible input data vectors **x**. The *expressibility* of an ensemble of unitaries is defined as how close the ensemble *uniformly* covers the unitary group. Specifically, one can measure the expressibility of a given ensemble by how close it is from a *2-design* (a pseudo-random distribution that agrees with the random distribution up to the second moment).

The *quantum kernel k(x, x')* requires computing the inner product between two vectors in an exponentially large Hilbert space. As such, for highly expressive encodings we are essentially evaluating the inner product between two approximately random (and hence orthogonal) vectors, thus leading to typical kernel values being exponentially small. That is, *higher encoding expressibility leads to greater quantum kernel concentration.*

Therefore, highly expressive encodings should be avoided. More specifically, unstructured encodings should be avoided and the *input data structure* should be taken into account when designing quantum feature encoding circuits, as is the case for covariant quantum kernels discussed in Section 4.1 Covariant Quantum Kernels.

## 5.2 Global-measurement-induced Concentration

A *global measurement* is a measurement that acts non-trivially on all n qubits. Such global measurements are required by design to compute quantum kernels. Global measurements can lead to *exponential concentration*, even when the expressibility of the quantum feature encoding circuit is low, because we are attempting to *extract global information* about a state that lives in an exponentially large Hilbert space.

Therefore, when using global measurements to evaluate the quantum kernel, the encoding must be chosen particularly carefully. To achieve this, one can take the *input data structure* into consideration when designing quantum feature encoding circuits, as before.

### 5.3 Noise-induced Concentration

*Hardware noise* may disrupt and destroy information in the encoded quantum states, providing a source of *exponential concentration*. The concentration of quantum kernels due to noise is exponential in the *number of layers L*, with the concentration stronger for *higher noise levels*.

Noise-induced concentration results pose a significant barrier to the successful implementation of quantum kernels on *NISQ devices*. Therefore, this is a warning *against using deep encoding schemes* in the NISQ era.

## 6 Conclusion

Quantum kernels with hybrid kernel methods, a.k.a. quantum kernel methods, offer distinct advantages as a hybrid approach to quantum machine learning. Not only do they apply to NISQ devices, but they also can be used to solve all types of machine learning problems including regression, classification, clustering, and dimension reduction. Furthermore, beyond quantum utility, quantum advantage can be attained if the quantum kernels, i.e., the quantum feature encodings, are classically intractable.

**Acknowledgement:** Thanks to my wife Hedy (郑期芳) for her support.